\documentclass[iop,apj,tighten,twocolappendix,numberedappendix]{emulateapj}
\usepackage{apjfonts}

\usepackage{graphicx}
\usepackage{amsmath}
\usepackage{amssymb}
\usepackage{color}
\usepackage{mathtools}
\usepackage[breaklinks,colorlinks,citecolor=blue,linkcolor=red]{hyperref} 
\usepackage[all]{hypcap}

\newcommand\msun{\, \rm M_\odot}

\begin{document} 

\title{A new method to constrain the origins of dark matter-free galaxies and their unusual globular clusters}
\author{Nathan W. C. Leigh\altaffilmark{1,2}, Giacomo Fragione\altaffilmark{3}}
\affil{$^1$Departamento de Astronom\'ia, Facultad de Ciencias F\'isicas y Matem\'aticas,
Universidad de Concepci\'on, Concepci\'on, Chile}
\affil{$^2$Department of Astrophysics, American Museum of Natural History, Central Park West and 79th Street, New York, NY 10024}
\affil{$^3$Racah Institute for Physics, The Hebrew University, Jerusalem 91904, Israel}

\begin{abstract}
We present a novel method to constrain the past collisional evolution of observed globular cluster (GC) systems, in particular their mass functions.  We apply our method to a pair of galaxies hypothesized to have recently undergone an episode of violent relaxation due to a strong galaxy-galaxy interaction, namely NGC 1052-DF2 and NGC 1052-DF4.  We begin by exploring the observational evidence for a collisional origin for these two recently discovered ultra-diffuse galaxies observed in the NGC 1052 group, posited in the literature to be dark matter (DM)-free. We compute the timescales for infall to the central nucleus due to dynamical friction (DF) for the GCs in these galaxies, using the shortest of these times to constrain how long ago a galaxy-galaxy interaction could have occurred.  
We go on to quantify the initial GC numbers and densities needed for significant collisional evolution to occur within the allotted times, and show that, if the hypothesis of a previous galaxy-galaxy interaction is correct, a paucity of low-mass GCs should be revealed by deeper observational surveys.  If any are found, they should be more spatially extended than the currently observed GC population. Finally, we apply our method to these galaxies, in order to illustrate its efficacy in constraining their dynamical evolution.  Our results motivate more complete observations of the GC luminosity functions in these galaxies, in addition to future studies aimed at combining the method presented here with a suite of numerical simulations in order to further constrain the origins of the curious GC populations in these (and other) galaxies. 
\end{abstract}


\section{Introduction}
\label{intro}

Recently, \citet{vandokkum18a} reported the discovery of a dark matter (DM)-free galaxy, namely the ultra-diffuse galaxy NGC 1052-DF2.  This galaxy is one of 23 objects identified in the group NGC 1052 using the Dragonfly Telescope Array \citep{abraham14,behroozi13}, and subsequently followed up using the ACS on the Hubble Space Telescope (HST) \citep{cohen18}.  The authors used the radial velocities of ten globular clusters (GCs) orbiting within the potential of this galaxy to constrain its velocity dispersion to be $\sim$ 10 km s$^{-1}$ \citep{vandokkum18b}.  They report a total luminous mass of 2 $\times$ 10$^8$ M$_{\odot}$ and, from its velocity dispersion, a total mass (seen and unseen) of 3.4 $\times$ 10$^8$ M$_{\odot}$.  This implies a ratio for M$_{\rm halo}$/M$_{\rm stars}$ of order unity, where M$_{\rm stars}$ is the total stellar mass and M$_{\rm halo}$ is the total galaxy mass including the DM halo.  Thus, the observations are consistent with there being no DM in this galaxy, since this ratio is typically at least a factor of $\sim 400$ higher \citep{behroozi13}.  The authors infer from this that DM is not always coupled to baryonic matter on galactic scales. 

In a subsequent paper, a second DM-free galaxy was reported.  NGC 1052-DF4 is a low surface brightness galaxy in the same group, identified by \citet{vandokkum19}.  The authors infer a total enclosed mass within 7 kpc of 0.4$^{+1.2}_{-0.3}$ $\times$ 10$^8$ M$_{\odot}$, and a total stellar mass of (1.5 $\pm$ 0.4) $\times$ 10$^8$ M$_{\odot}$ within the same enclosed radius.  They conclude that this galaxy is consistent with having no DM.  As with NGC 1052-DF2, this galaxy hosts an unusually bright population of GCs, but more extended than NGC 1052-DF2.

The existence of such dark-matter deficient galaxies is still disputed, however. For example, \citet{trujillo19} recently argued that the distance to NGC 1052 is only 13 Mpc instead of the 20 Mpc measured by \citet{vandokkum18a}.  The authors further argue that this can explain \textit{both} the proposed lack of DM and the anomalous GC populations.  With that said, \citet{vandokkum18c} subsequently showed that the colour-magnitude diagram is strongly influenced by blends, causing the appearance of a false red giant branch tip about roughly twice as bright as the true red giant branch tip.  This translates into an underestimate of the true distance by a factor of $\sim$ 1.4.  \citet{laporte19} further argue that an underestimate of the uncertainty on the mass of the host galaxy could also explain the need to invoke DM-free halos in these host galaxies.  As an independent explanation for the apparently curious observational results of \citet{vandokkum18a}, \citet{kroupa19} proposed that the apparent lack of DM in these galaxies can be understood within the context of MOdified Newtonian Dynamics (MOND), which should cause a weaker self-gravity in the outskirts of galaxies when in close proximity to a massive host.  In spite of these interesting counter-arguments to the work of \citet{vandokkum18a}, these works do not explain the curious GC luminosity function and spatial distribution, at least not without a paucity of low-mass GCs relative to the Milky Way and other galaxies (see Figure~\ref{fig:fig1} below). 

How might a DM-free galaxy form?  One possibility relies on impulsive heating mediated by tidal forces.  This can in principle alter a rotationally-supported disk of stars and gas into a spheroidal structure.  Often termed tidal stripping or shocking \citep[e.g.][]{gnedin99,mayer07}, this mechanism may require an additional process to fully deplete the new spheroid of its gas \citep[e.g.][]{maclow99}.  \citet{donghia09} considered direct interactions between dwarf disk galaxies and more massive interlopers.  Using numerical simulations, the authors describe a mechanism they term "resonant stripping" that can strip dwarf disk galaxies of their stars.  The mechanism occurs for prograde encounters with large mass ratios of order $\sim 10$--$100$. Resonant stripping happens when the spin and orbital frequencies are comparable.  This pulls the gas and stars out of the galaxy, since they comprise the disk, whereas the DM is not affected since it is pressure-supported and has no spin frequency. 

Several authors have pointed out that the observed populations of GCs in NGC 1052-DF2 and NGC 1052-DF4 are both peculiar \citep{emse19,fens19}.  In particular, where are all the low-mass GCs?  And why are the observed GCs so centrally concentrated?  These galaxies lie well off the previously reported relation between the total GC mass in galaxies and the total mass of their DM halos \citep{choksi18}.  Figure~\ref{fig:fig1} shows a comparison between these two GC populations and the Milky Way (MW) GC population.  First, even though the MW is much more massive and also more extended than either NGC1052-DF2 or NGC1054-DF4, we see that the latter galaxies have a larger fraction of very bright/massive GCs at small Galactocentric radii when compared to the MW's GCs.  As shown by the open squares and dotted histograms, this remains the case, although to a lesser extent, if one adopts the distance estimate provided in \citet{trujillo19}.  Second, the nearest giant elliptical galaxy to the MW, namely NGC 5128 (i.e., Centaurus A), is home to a population of GCs whose mass function is similar to that of the M31 GC system but with a larger mean GC mass (and also mean mass-to-light ratio), and indistinguishable from the MW's GC system (due mostly to the much smaller sample size in the MW compared to NGC 5128) \citep[e.g.][]{taylor15} .  The GC populations in NGC 1052-DF2 and NGC 1052-DF4 are therefore probable outliers in previously reported studies looking at, for example, GC mass functions and GC specific frequencies in different types of galaxies \citep[e.g.][]{harris13,harris16}.  Additionally, the GCs in NGC1052-DF2 and NGC1052-DF4 are significantly more concentrated at small (projected) galactocentric distances compared to the brightest GCs in the sample from \citet{harris96}. 

\begin{figure}
\begin{center}
\includegraphics[width=\columnwidth,height=9cm]{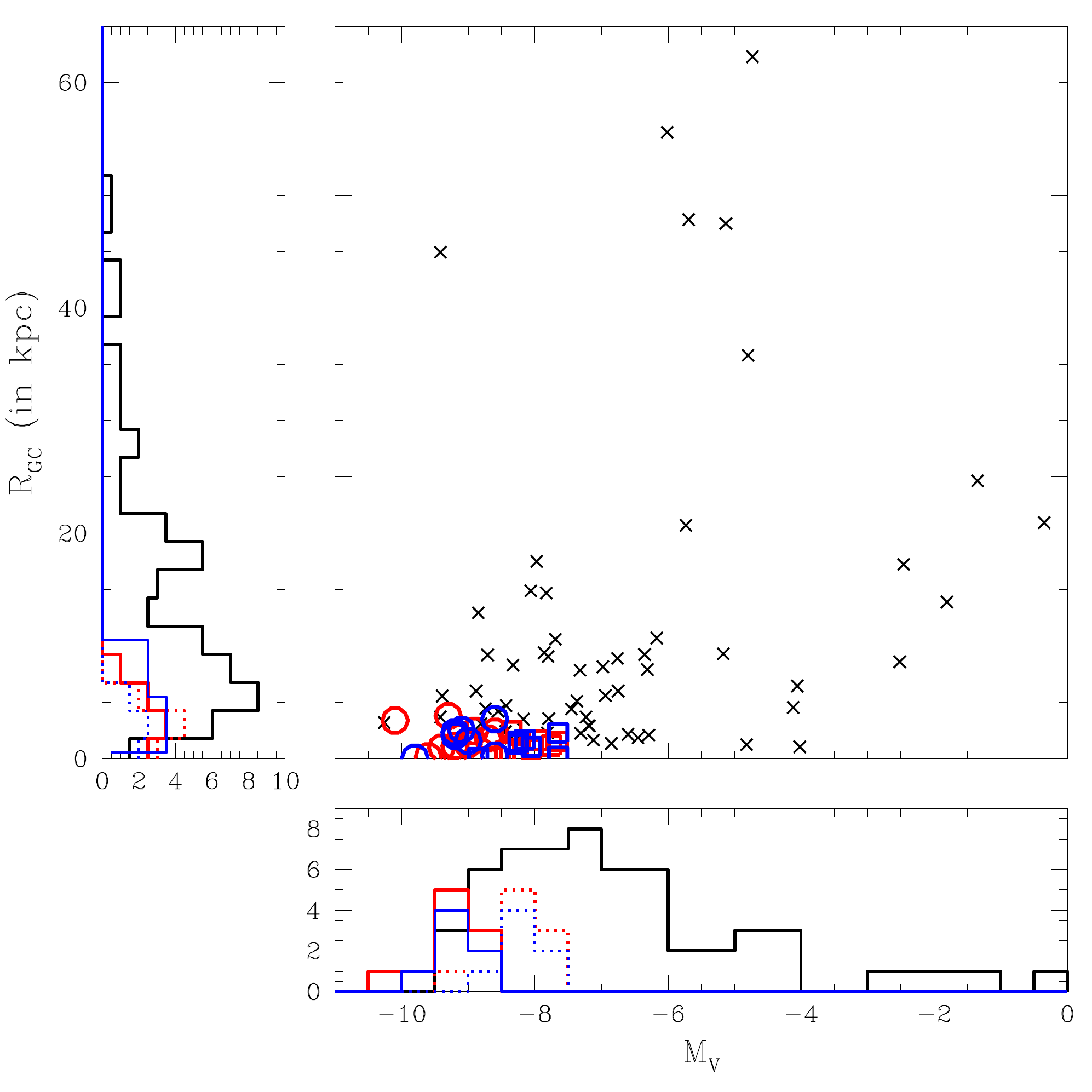}
\end{center}
\caption[Comparison between Milky Way GCs and those GCs in NGC1052-DF2 and NGC1052-DF4]{The integrated V-band magnitudes for the Milky Way GC population (shown by the black crosses) are plotted against their Galactocentric distances, using all available data in \citet{harris96}.  For comparison, we also plot the same observed quantities  for the GCs in NGC1052-DF2 (red open circles) and NGC1052-DF4 (blue open circles) using the distance of 20 Mpc assumed in \citet{vandokkum18a}.  The open red and blue squares, as well as the dotted histograms, show the same thing but adopting the distance estimate of 13 Mpc found by \citet{trujillo19}.  Note that we have not included the new GC candidates reported in \citet{trujillo19}, since this comparison has already been done in their Figures 11 and 12.   
\label{fig:fig1}}
\end{figure}

In this paper, we consider a scenario in which the ultra-diffuse galaxies NGC 1052-DF4 and NGC 1052-DF2, argued by some in the literature to be DM-free, were dynamically stripped of their DM haloes.  This could have occurred, for example, due to a strong interaction with a more massive nearby galaxy \citep[e.g.][]{donghia09,ogiya18} or even a direct galaxy-galaxy collision \citep[e.g.][]{silk19}, which triggered an episode of violent relaxation in their GC populations \citep{lyndenbell67}. We then consider the subsequent dynamical evolution of such a disturbed GC population.  

We begin by motivating the need for consideration of the above scenario.  First, we compute numerically the mass, density and velocity dispersion profiles of both NGC 1052-DF2 and NGC 1052-DF4.  These are used to compute the radial profiles of the DF timescales in both galaxies for a typical GC with a mass of 10$^6$ M$_{\odot}$.  Second, we calculate the dynamical friction timescales for all GCs reported in \citet{vandokkum18b} and \citet{vandokkum19}, and compare them to a Hubble time. If a computed DF timescale is much shorter than a Hubble time, we interpret this as evidence that they did not form in their currently observed positions, motivating consideration of other formation scenarios.  

As we will show, the analysis described above reveals one and two candidate GCs in, respectively, NGC 1052-DF2 and NGC 1052-DF4 with unusually short DF timescales.  This in turn motivates the development of methods that can used to constrain the origins of such galaxies with curious empirical properties.  In this paper, we are most interested in explaining the properties of their globular cluster populations, which could hint at a significantly perturbed dynamical evolution.  Ideally, such methods can then be applied to the available observational data to constrain the origins of these galaxies and their curious GC populations, and/or make predictions for future data sets.   We further present a novel method to constrain the collisional evolution for such GC populations post-galaxy-galaxy interaction.  We argue using our analytic method that numerical simulations combined with observed constraints on the GC luminosity function can be used to constrain the GC mass function immediately post-interaction, motivating the need for such future theoretical and observational studies. 

Our methods and results are presented in Section~\ref{method}. In Section~\ref{discussion}, we discuss the implications of our results for understanding the origins of the hypothesized DM-free galaxies and their GC populations, and make predictions for the observed properties of future discoveries in this potentially new class of galaxies.  Finally, we conclude in Section~\ref{summary}.

\section{Calculations}
\label{method}

In this section, we compute numerically the mass, density and velocity dispersion profiles of both NGC 1052-DF2 and NGC 1052-DF4.  These are used to compute order-of-magnitude estimates for the radial profiles of the DF timescales in both galaxies for a typical GC with a mass of 10$^6$ M$_{\odot}$.  We also calculate the dynamical friction timescales for all GCs reported in \citet{vandokkum18b} and \citet{vandokkum19} and compare these to a Hubble time.

\subsection{Mass, density and velocity dispersion profiles}

In order to calculate the density and velocity dispersion profiles, we must first calculate the mass enclosed within radius $r$. This requires obtaining the free parameters in the fitting functions from previous observational studies focused on the two galaxies in our sample.  \citet{cohen18} fit the surface brightness profile of the dwarf spheroidal galaxy NGC 1052-DF4 using a Sersic model. We follow these authors and adopt a Sersic index of n $=$ 0.79, a central surface brightness of $\mu$(V$_{\rm 606,0}) =$ 23.7 and a major axis half-light radius of R$_{\rm e} =$ 1.6 kpc, and assume a distance to the galaxy of D $=$ 20 Mpc.  

We perform an analogous calculation for NFC 1052-DF2.  \citet{cohen18} also fit the surface brightness profile of the dwarf spheroidal galaxy NGC 1052-DF2 using a Sersic model.  We adopt the same parameters as these authors, specifically a Sersic index of n $=$ 0.55, a central surface brightness of $\mu$(V$_{\rm 606,0}) =$ 24.2, a (major-axis) half-light radius of R$_{\rm e} =$ 1.8 kpc, and a distance to the galaxy of D $=$ 20 Mpc.  

To calculate the mass profiles for both galaxies, we adopt Equation A2 for the enclosed mass M(r) from \citet{terzic05}:
\begin{equation}
\label{eqn:mass}
M(r) = 4{\pi}{\rho_{\rm 0}}R_{\rm e}^3nb^{n(p-3)}\gamma(n(3-p),z),
\end{equation}
where $\gamma$ is the incomplete gamma function, and the dimensionless variable is defined as:
\begin{equation}
\label{eqn:zedd}
z = b\Big( \frac{r}{R_{\rm e}} \Big)^{1/n},
\end{equation}
and, after a little math:
\begin{equation}
\label{eqn:rho0}
\rho_{\rm 0} = \frac{\sqrt{{\pi}}}{4R_{\rm e}}{\Upsilon_{\rm 0}}I_{\rm e}b^{n(1-p)}.
\end{equation}
Finally, the mass-to-light ratio for an old stellar population is typically $\Upsilon_{\rm 0} = M/L \sim$ 2 $\msun/\mathrm{L}_\odot$, and the variable p can be approximated by the relation:
\begin{equation}
\label{eqn:pparam}
p = 1 - \frac{0.6097}{n} + \frac{0.055}{n^2}.
\end{equation}
We adopt n $=$ 0.79, for which Equation~\ref{eqn:pparam} reduces to p $=$ 0.3163. Finally, the radial velocity dispersion profile is computed using Equation A5 in \citet{terzic05}.  Note that we assume that our target galaxies are DM-free in the preceding calculations.

\subsection{Calculating dynamical friction timescales} \label{DFtimes}

The term dynamical friction, in its original form, refers to the gravitational focusing of particles into a wake by a massive perturber as it travels through a homogeneous background medium of constant density \citep{chandrasekhar43}.  As applied to GCs orbiting in the potentials of their host galaxy \citep{tremaine75}, this generates a damping force due to the gravitational tug of the trailing wake, and ultimately removes energy and angular momentum from the GC's orbit, causing it to (eventually) spiral into the host galaxy's centre of mass. Assuming circular orbits, the timescale for dynamical friction to operate is approximately given by \citep{binney87,gnedin14}:
\begin{equation}
\label{eqn:tau_df}
\tau_{\rm df} = \frac{1.17M(r)r}{\ln{\Lambda}m_{\rm GC}\sigma(r)},
\end{equation}
where M(r) and $\sigma$(r) are, respectively, the enclosed galaxy mass and the stellar velocity dispersion at a distance $r$ from the centre of mass of the galaxy, m$_{\rm GC}$ is the mass of the orbiting GC and $\ln{\Lambda}$ is the Coulomb logarithm for which we adopt $\ln{\Lambda} = 10$ \citep[for details see][]{arcas17,nusser19}.

\subsection{Radial profiles} \label{radial}

\begin{figure}
\begin{center}
\includegraphics[width=\columnwidth,height=10cm]{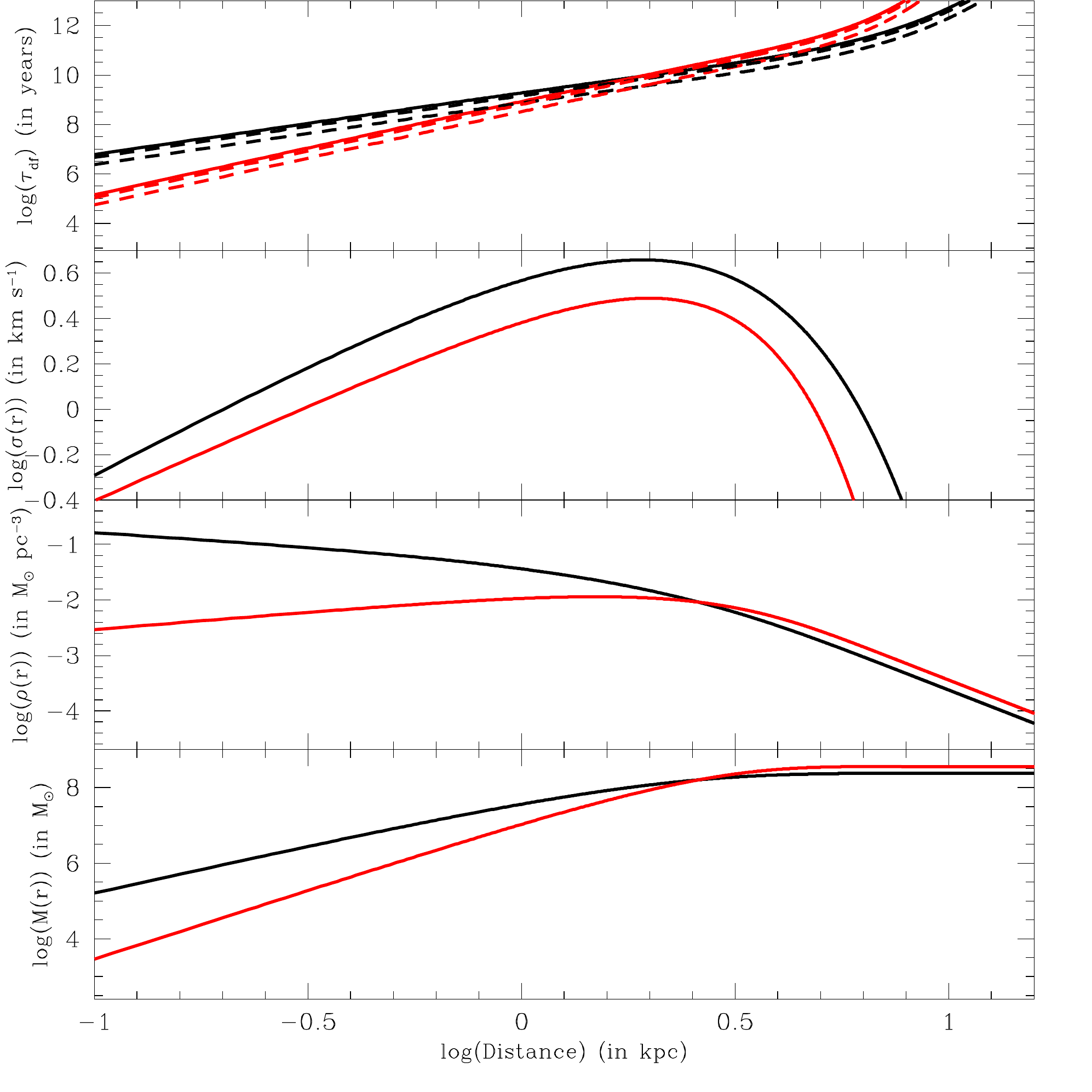}
\end{center}
\caption[The radial dependences of the enclosed stellar mass, mass density, velocity dispersion and DF timescale of both NGC 1052-DF4 and NGC 1052-DF2.]{From bottom to top, each panel shows the radial dependence of, respectively, the enclosed mass M(r), the mass density $\rho$(r), the stellar velocity dispersion $\sigma$(r) and the DF timescale $\tau_{\rm df}$ for a hypothetical GC with total mass m$_{\rm GC} =$ 10$^6$ M$_{\odot}$ and assuming for the Coulomb logarithm $\ln{\Lambda} =$ 6, for the galaxies NGC 1052-DF4 (black) and NGC 1052-DF2 (red).  The dashed lines in the top panel show how the DF timescales are expected to change assuming eccentric orbits, adopting the minimum and maximum correction factors provided in \citet{gnedin14} (see text for more details). 
\label{fig:fig2}}
\end{figure}

The radial dependences of the host galaxy enclosed masses, densities and velocity dispersions are shown in Figure~\ref{fig:fig2}.  We also show the DF timescales as a function of distance from the centre of mass of the host galaxy using Equation~\ref{eqn:tau_df}, for a hypothetical GC with total mass m$_{\rm GC} =$ 10$^6$ M$_{\odot}$.  The dashed lines in the top panel show the corrected DF timescales assuming eccentric orbits.  Specifically, we multiply the DF timescales by the minimum and maximum correction factors provided in \citet{gnedin14}, which are 0.4 and 0.8, respectively (see Section~\ref{DF}). \textit{More eccentric orbits reduce the DF timescale at a given galactocentric distance.} 

We consider an isotropic model in deriving our velocity dispersion profiles, shown in the third inset of Figure~\ref{fig:fig2}. The velocity dispersion peaks very close to the observed projected galactocentric distances of many GCs in both NGC 1052-DF2 and NGC 1052-DF4, and declines rapidly on either side of this peak.  We caution, however, that the peaks of our velocity distributions, in particular for NGC 1052-DF2, are slightly lower than that reported in \citet{vandokkum18b}.  This could translate into mildly higher DF timescales for this galaxy than expected from the velocity measurements of \citet{vandokkum18b}.  This should be accounted for when trying to infer the \textit{true} DF timescales.  However, these analytic estimates are at best approximations (see the next section).  In spite of this, our basic conclusions are consistent with those of \citet{chowdhury19} and \citet{nusser19}.  

\begin{table}
\caption{The computed properties for all globular clusters in both NGC 1052-DF2 (top 10 rows) and NGC 1052-DF4 (bottom 7 rows). The last two columns show the computed DF timescales for each GC, assuming a distance to the NGC 1052 group of 20 Mpc (column 4) and 13 Mpc (column 5).}
\centering
\begin{tabular}{ccccc}
\hline
GC ID      &  Mass  &    Distance (kpc)  &  $\tau_{\rm df}$ (20 Mpc) & $\tau_{\rm df}$ (13 Mpc) \\
                &    (M$_{\odot}$)   & (kpc) & (Gyr) & (Gyr) \\ 
\hline
39  &  6.7 $\times$ 10$^5$   &  7.55  &  9800  &  39000 \\
59  &  4.7 $\times$ 10$^5$   &  4.91  &  720  & 2900 \\
71  &  5.1 $\times$ 10$^5$   &  2.57  &  52  & 210 \\
73  &  1.4 $\times$ 10$^6$   &  6.77  &  1800  & 7100 \\
77  &  8.9 $\times$ 10$^5$   &  7.55  &  7400  & 30000 \\ 
85  &  6.2 $\times$ 10$^5$   &  2.26  &  28 & 110 \\
91  &  6.2 $\times$ 10$^5$   &  1.55  &  7.1  & 28 \\
92  &  7.4 $\times$ 10$^5$   &  1.94  &  13 &  51  \\
98  &  3.9 $\times$ 10$^5$   &  3.59  &  230 &  940 \\
101 &  3.5 $\times$ 10$^5$   &  4.77  &  830  & 3300 \\
\hline
2726  &  6.2 $\times$ 10$^5$   &  4.69  &  150 & 620  \\
2537  &  6.2 $\times$ 10$^5$   &  4.08  &  100  & 400 \\
2239  &  3.5 $\times$ 10$^5$   &  0.57  &  1.4  & 5.6 \\
1968  &  1.1 $\times$ 10$^6$   &  2.90  &  23  & 92 \\
1790  &  5.1 $\times$ 10$^5$   &  3.17  &  60  & 240 \\ 
1452  &  5.6 $\times$ 10$^5$   &  5.13  &  230 & 940 \\
943   &  3.5 $\times$ 10$^5$   &  7.01  &  1400  & 5800 \\
\hline
\end{tabular}
\label{table:one}
\end{table}

\subsection{Dynamical friction timescales for individual GCs}
\label{DF}

In this section, we compute DF timescales for all seven and ten GCs orbiting within, respectively, the galaxies NGC 1052-DF4 and NGC 1052-DF2 reported in \citet{vandokkum19} and \citet{vandokkum18b}.  

\subsubsection{Timescales}

In this section, we compute DF timescales for all seven and ten GCs orbiting within, respectively, the galaxies NGC 1052-DF4 and NGC 1052-DF2 reported in \citet{vandokkum19} and \citet{vandokkum18b}.  

\begin{figure}
\begin{center}
\includegraphics[width=\columnwidth]{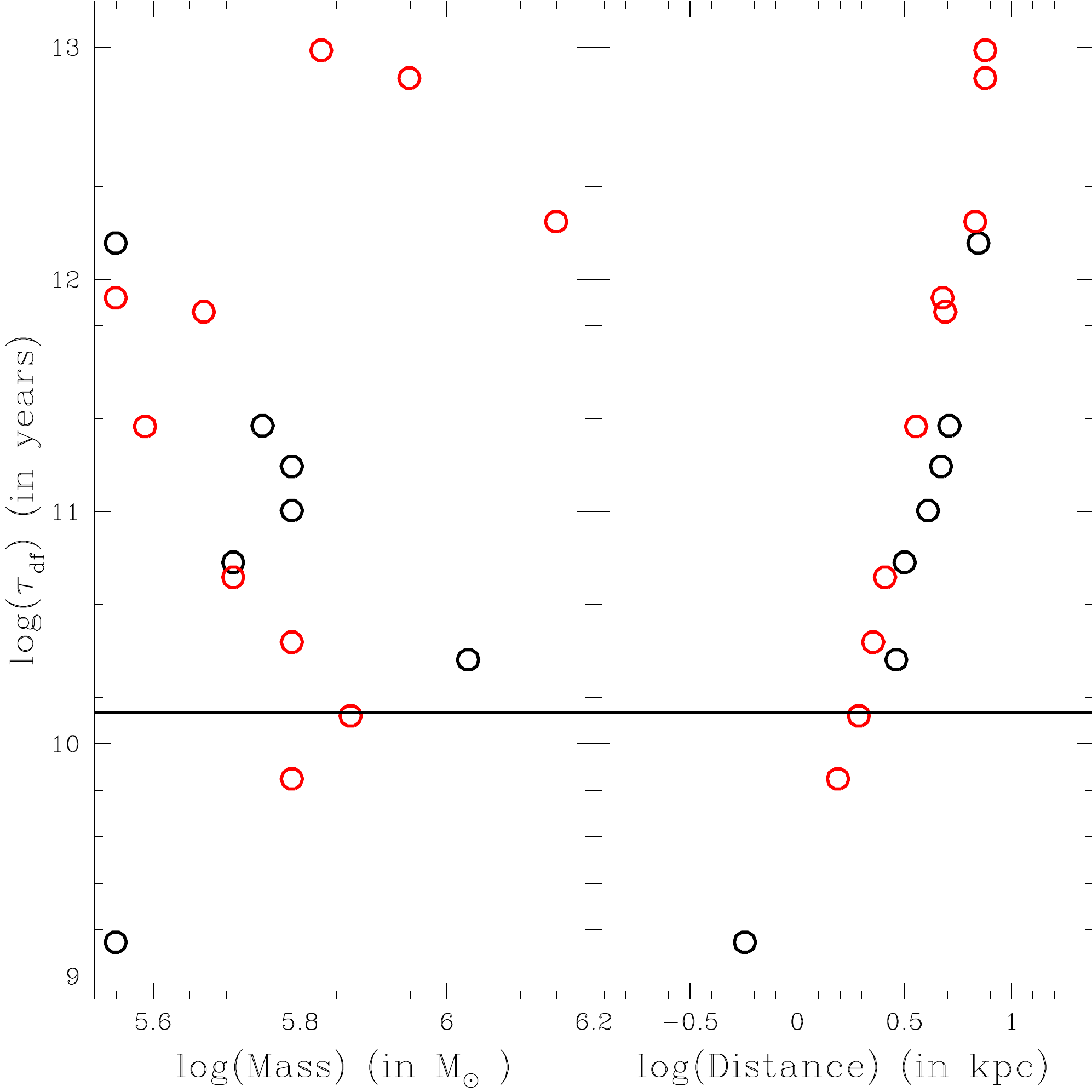}
\end{center}
\caption[DF timescales for all 7 GCs observed in NGC 1052-DF4, and all 10 GCs observed in NGC 1052-DF2]{The DF timescales are shown as a function of the total GC mass (left panel) assuming a mass-to-light ratio of 2 $\msun/\mathrm{L}_\odot$ for all GCs, and as a function of the projected galactocentric distance (right panel) in kpc.  These results are shown for all 7 GCs in NGC 1052-DF4 (black), and all 10 GCs in NGC 1052-DF2 (red).  The horizontal solid line demarcates a Hubble time. 
\label{fig:fig3}}
\end{figure}

Using Equations~\ref{eqn:tau_df} and Equation A5 from \citet{terzic05}, we show the computed DF timescales in Figure~\ref{fig:fig3} and Table~\ref{table:one}, assuming a distance to NGC 1052-DF4 and NGC 1052-DF2 of 20 Mpc.  The left panel shows the DF timescales as a function of the total GC mass, assuming a mass-to-light ratio of 2  $\msun/\mathrm{L}_\odot$, whereas the right panel shows the same timescales but as a function of the projected galactocentric distance.  The horizontal solid line demarcates a Hubble time.  One GC has a DF timescale shorter than a Hubble time in NGC 1052-DF4 (black open circles), whereas two out of ten GCs have DF timescales less than a Hubble Time in NGC 1052-DF2 (red open circles).  

\subsubsection{Uncertainties and assumptions}

We caution that our computed DF timescales should come along with significant uncertainty, stemming mostly from the GCs' \text{true or 3D} distances from their host galaxy centre of mass, which are not known and could be larger than their observed \textit{projected} galactocentric distances (although, as discussed in \citet{nusser19}, this simple analytic calculation also suffers from issues related to, for example, a more complicated galaxy mass profile than is represented by our analytic approximations, interactions between GCs, etc.).

Is it possible that the eccentricities of the observed GCs are non-negligible? If so, this could yield DF timescales shorter than we find by assuming that they are on circular orbits. To include the effect of the eccentricity, we consider a reduced DF timescale:
\begin{equation}
\tau_{\rm df,ecc} = \tau_{\rm df}\left(\frac{J}{J_c(E)}\right)^\alpha\ ,
\label{tau_dfecc}
\end{equation}
where $J/J_c(E)$ is the ratio of the orbital angular momentum to its maximum value for a given energy $E$. The values of the exponent given in the literature range from $\alpha\approx 0.4$ \citep{colpi99} to $\alpha\approx 0.8$ \citep{lacey93}.

To quantify this effect, we use a Monte Carlo approach. In our calculation, we assume that the ratio $J/J_c(E)$ is uniformly distributed for the GC population under consideration. We sample $J/J_c(E)$ for 10$^4$ realizations and compute the typical DF timescale in Eq.~\ref{tau_dfecc}. We find that the number of GCs expected to have $\tau_{\rm df,ecc} <$ 10$^{10}$ yr remains constant both for NGC 1052-DF2 and NGC 1052-DF4. Thus, the (unknown) eccentricities of the observed GCs do not significantly affect their DF timescales.

\begin{table}
\caption{Fraction of clusters (out of the total) with $\tau_{\rm df}<10^{10}$ yr for different assumptions for GC eccentricity and projected distance (see Section~\ref{DF}).}
\centering
\begin{tabular}{cccc}
\hline
Galaxy      &  Eccentricity   &    Distance  &  Fraction \\
\hline
NGC 1052-DF2   &  circular                 &  $r_{\rm obs}$            &  $10\%$ \\
NGC 1052-DF2   &  $({J}/{J_c(E)})^{0.4}$   &  $r_{\rm obs}$            &  $10\%$ \\
NGC 1052-DF2   &  $({J}/{J_c(E)})^{0.8}$   &  $r_{\rm obs}$            &  $10\%$ \\
NGC 1052-DF2   &  circular                 &  $r_{\rm obs}/\cos\theta$  &  $5.6\%$ \\
NGC 1052-DF4   &  circular                 &  $r_{\rm obs}$            &  $29\%$ \\
NGC 1052-DF4   &  $({J}/{J_c(E)})^{0.4}$   &  $r_{\rm obs}$            &  $29\%$ \\
NGC 1052-DF4   &  $({J}/{J_c(E)})^{0.8}$   &  $r_{\rm obs}$            &  $29\%$ \\
NGC 1052-DF4   &  circular                 &  $r_{\rm obs}/\cos\theta$  &  $13\%$ \\
\hline
\end{tabular}
\label{table:two}
\end{table}

Next, we attempt to quantify the possible importance of projection effects. Specifically, could any of the GCs (especially the three with $\tau_{\rm df} <$ $\tau_{\rm Hubble}$ in Figure~\ref{fig:fig3}) have true 3D galactocentric distances much larger than their observed \textit{projected} distances? If so, this is important, since it would give rise to an artificially short DF timescale for some GCs. To address this possibility, we assume that the true distance $r_{\rm true}$ from the centre of mass of the host galaxy is related to the observed distance $r_{\rm obs}$ (see Table~\ref{table:two}) by:
\begin{equation}
r_{\rm obs}=r_{\rm true}\cos \theta\ ,
\end{equation}
where $\theta$ is the angle between the true vector and the projected vector. We then sample $\cos \theta$ uniformly for $10^4$ realizations, compute $r_{\rm true}$ and $\tau_{\rm df}$ via Eq.~\ref{eqn:tau_df}. In the case of NGC 1052-DF2, we find that the probability for GC 77 to still have $\tau_{\rm df}$ less than a Hubble time is $\sim 56\%$. In the case of NGC 1052-DF4, the probability to have two or one GCs with $\tau_{\rm df}$ less than a Hubble time is $\sim 43\%$ and $\sim 47\%$, respectively. 

\textit{Thus, neither non-zero eccentricities for the observed GCs or projection affects should significantly affect any of our conclusions thus far.}

Finally, what if the distances to NGC 1052-DF2 and NGC 1052-DF4 are wrong, as suggested in \citet{trujillo19}?  These authors argue for a distance of only 13 Mpc instead of the 20 Mpc adopted in \citet{vandokkum18a}.  We have computed the dynamical friction timescales under the assumption that both NGC 1052-DF2 and NGC 1052-DF4 are DM-free. However, assuming a distance of only 13 Mpc, \citet{trujillo19} argued that the minimum DM mass would be $\sim 10^9\msun$, with a stellar mass of $\sim 10^7\msun$, for NGC 1052-DF2. If we account for this in Eq.~\ref{eqn:tau_df}, along with the different GC V-band magnitudes and galactocentric distances (see Fig.~\ref{fig:fig1}), $\tau_{\rm df}$ would be a factor of $\sim 4$ longer than if these galaxies are DM-free. Hence, to correct the DF timescales in Table~\ref{table:one} for these new distances, we simply multiply the DF timescales in column 4 by this correction factor. Thus, only the GC 2239 in NGC 1052-DF4 would have a DF timescale shorter than a Hubble time. 

If the distances reported in \citet{trujillo19} are correct, then this could weaken any DF-based arguments discussed in this paper.  With that said, it is important to keep in mind that the computed DF timescales do not account for GC-GC interactions, which could impede their infall toward the host galaxy nucleus.  As quantified in subsequent sections, one of the purposes of this paper is to address the possibility of such GC-GC interactions.  Furthermore, the distance proposed by \citet{trujillo19} would not completely explain the unusual GC luminosity functions in these galaxies.  As argued by these authors, it would shift the GC luminosity function to lower luminosities, but would still not explain the fact that they appear centrally concentrated, nor would it explain an apparent (but unconfirmed) paucity of lower mass GCs.

\subsubsection{What are the computed DF timescales telling us about the origins of these GCs?}

As reported above, we find that one out of seven GCs has a DF timescale shorter than a Hubble time in NGC 1052-DF4, whereas two out of ten GCs have DF timescales less than a Hubble Time in NGC 1052-DF2.  But what is this telling us about their origins?

If the computed DF timescales in Table~\ref{table:one} are taken at face value, this suggests that even if these GCs began further out in their host galaxy potentials and migrated in to their currently observed galactocentric distances, we have been fortuitous to have caught \textit{both} GC 2239 in NGC 1052-DF4 \textit{and} GC 91 in NGC 1052-DF2 just before inspiral in to the nucleus.  Independent of the issue of dynamical friction, it is remarkable that the GC system mass is a few percent of the galaxy stellar mass in both galaxies, whatever distance is assumed (see the discussion in the next section). These galaxies remain peculiar in the well-known cosmological scaling relations, compared to, for example, dwarf galaxies which have similar stellar masses \citep[see e.g.][]{gnedin14,forbes18}. Thus, some mechanism must have resulted in NGC 1052-DF2 and NGC 1052-DF4 having a large fraction of their baryonic mass in the form of massive luminous GCs.

Could the above curiosities be explained by significant collisional evolution of these GC populations in the past, perhaps triggered by a previous galaxy-galaxy interaction?  Such an interaction would have most likely contributed to a phase of violent relaxation, and in so doing could have initiated a potentially rapid subsequent dynamical evolution in the collisional regime.  This would most likely be followed by a gentler re-distribution of orbital energies toward re-entering a state of approximate equipartition of the GC populations, roughly operating on a relaxation timescale which is generally comparable to the DF timescale.

In the subsequent sections, we further constrain the dynamical origins of these GC populations, by asking if strong (i.e., with significant energy exchange) close interactions and/or direct GC-GC collisions could have realistically occurred.  Such collisions could have contributed to the unusual observed GC luminosity functions in these galaxies, by skewing them to larger GC masses.  We then proceed to present our method for constraining the dynamical histories of observed GC populations.

\subsection{The rate of direct GC-GC collisions} \label{coll}

Given the calculations presented in the previous sections, we expect to find that some GCs in NGC 1052-DF2 and NGC 1052-DF4 reside at or very near the centers of these galaxies, perhaps appearing as nuclear star clusters.  Indeed, given the low masses of these galaxies, known scaling relations predict that nuclear clusters should be present, instead of super-massive black holes \citep[e.g.][]{leigh12,leigh15,neumayer20}, yet they are not observed.  As suggested by \citet{chowdhury19}, GC-GC interactions could help to impede dynamical friction, continually stirring the centrally concentrated GC population and preventing them from falling in to the very centre of their host galaxy.  Both \citet{nusser19} and \citet{chowdhury19} find in their N-body simulations that such strong GC-GC interactions do occur frequently and that this could indeed contribute to slowing the rate of DF.  Hence, given that strong GC-GC interactions could suppress dynamical friction and prevent GC infall, consideration of such direct GC-GC interactions and even collisions could potentially help to solve the aforementioned problems related to the computed DF timescales and the lack of observed nuclear star clusters. 

To address this question, we first compute the mean times corresponding to direct collisions between GCs.  
\begin{eqnarray}
\label{eqn:coll}    
\tau_{\rm coll} &=& 1.1 \times 10^{10} \Big( \frac{1\ {\rm pc}}{R_{\rm GC,max}} \Big)^3\Big( \frac{10^3}{n_{\rm GC}} \Big)^2\times \nonumber\\
&\times& \Big( \frac{v_{\rm rms}}{5\ {\rm km s}^{-1}} \Big) \Big( \frac{0.5\ {\rm M}_{\odot}}{\bar{m}_{\rm GC}} \Big) \Big( \frac{0.5\ {\rm R}_{\odot}}{\bar{r}_{\rm GC}} \Big)\ {\rm yr},
\end{eqnarray}
where $\bar{r_{\rm GC}}$ is the mean GC half-light radius, $\bar{m_{\rm GC}}$ is the mean GC mass, v$_{\rm rms}$ is the root-mean-square velocity of the GC system (i.e., $\sqrt{3}$ times the line-of-sight velocity dispersion), R$_{\rm GC,max}$ is the maximum projected galactocentric distance in the galaxy and n$_{\rm GC}$ is the GC number density inside this volume. 

Equation~\ref{eqn:coll} has been adapted from Equation A9 in \citet{leigh11}.\footnote{We set f$_b$ and f$_t$ both equal to zero as these are the binary and triple fractions, respectively, in a star cluster.  Hence, in the original equation, these terms correct for the fraction of the total number of stars that are isolated singles.  For the present problem, we assume that all GCs (the equivalent of stars here in the original equation) are isolated single objects.}  In particular, it has been adapted from a roughly constant-density cluster core\footnote{The assumption of a constant density inner core is reasonable over the small spatial extent of the host galaxy that we consider here, since the inner-most density profile is not cuspy \citep{vandokkum18b,vandokkum19}.  A more detailed correction accounting for the shape of the inner host galaxy potential would affect the calculated timescales by at most a factor of order unity.} such that for the size or volume of the region of interest, we adopt the maximum galactocentric distance observed for all GCs in each galaxy.\footnote{We set r$_{\rm c}$ in Equation A9 to 7.55 kpc in NGC 1052-DF2 and 7.01 kpc in NGC 1052-DF4; see Table~\ref{table:one}).}  It is then straight-forward to compute a GC number density n$_{\rm GC}$ for each galaxy within this volume, by adopting 7 and 10 GCs for, respectively, NGC 1052-DF4 and NGC 1052-DF2 for the total number of GCs inside this volume.  For the velocity dispersions of the GC populations within these volumes, we take the likelihood values of 3.2 km s$^{-1}$ and 3.8 km s$^{-1}$ for, respectively, NGC 1052-DF2 \citep{vandokkum18b} and NGC 1052-DF4 \citep{vandokkum19}.  Using the indicated GC masses in Table~\ref{table:one} we calculate an average GC mass for each galaxy, and adopt a typical GC size (i.e., r$_{\rm GC}$) of 20 pc for all galaxies (motivated by a measured mean GC half-light radius of 6.5 $\pm$ 0.5 pc in NGC 1052-DF2 by \citet{vandokkum18b}).  With these parameters, we compute mean GC-GC interaction times corresponding to direct collisions of 460 Gyr and 730 Gyr for, respectively, NGC 1052-DF2 and NGC 1052-DF4. Over a period of 10 Gyr, this implies collision probabilities of only a few percent.  

The above collision times can be regarded as strict upper limits.  There are several reasons for this.  First, we consider only \textit{direct} collisions, in which the radii of the GCs overlap directly.  Significant energy should be exchanged for larger impact parameters, however, which could increase the above collision probabilities by up to about an order of magnitude.  Given the low velocity dispersions and hence escape velocities in these galaxies, the tendency toward equipartition would then readily contribute to the ejection of preferentially low-mass GCs from their host galaxies. In turn, this would rob preferentially more massive GCs of orbital energy and angular momentum, causing them to sink deeper in their host galaxy potentials. This could help to account for the observed unusually high masses and small galactocentric radii in NGC 1052-DF2 and NGC 1052-DF4, relative to the MW GC population (see Figure~\ref{fig:fig1}).  A similar result was also recently found by \citet{madau19}, who considered galaxy-galaxy collisions.  The authors pointed out that dissipative effects during the interactions (e.g., tides) should further contribute to more centrally concentrated GC populations.  

Second, the above simple calculations neglect any previous dynamical evolution of the GC populations - i.e., it assumes that what we see now for the GC populations is what has always been there.  For example, if the number density of GCs had been higher by a factor of 10 in the past \citep[see e.g.][]{frag18a,frag18b}, then the interaction rates would increase by a factor of 100 via Equation~\ref{eqn:coll}.  This would have resulted in a number of direct GC-GC collisions within a 10 Gyr period, while also potentially ejecting even more (preferentially low-mass) GCs from their host galaxy due to strong interactions and the tendency toward energy equipartition.

\begin{figure}
\begin{center}
\includegraphics[width=\columnwidth]{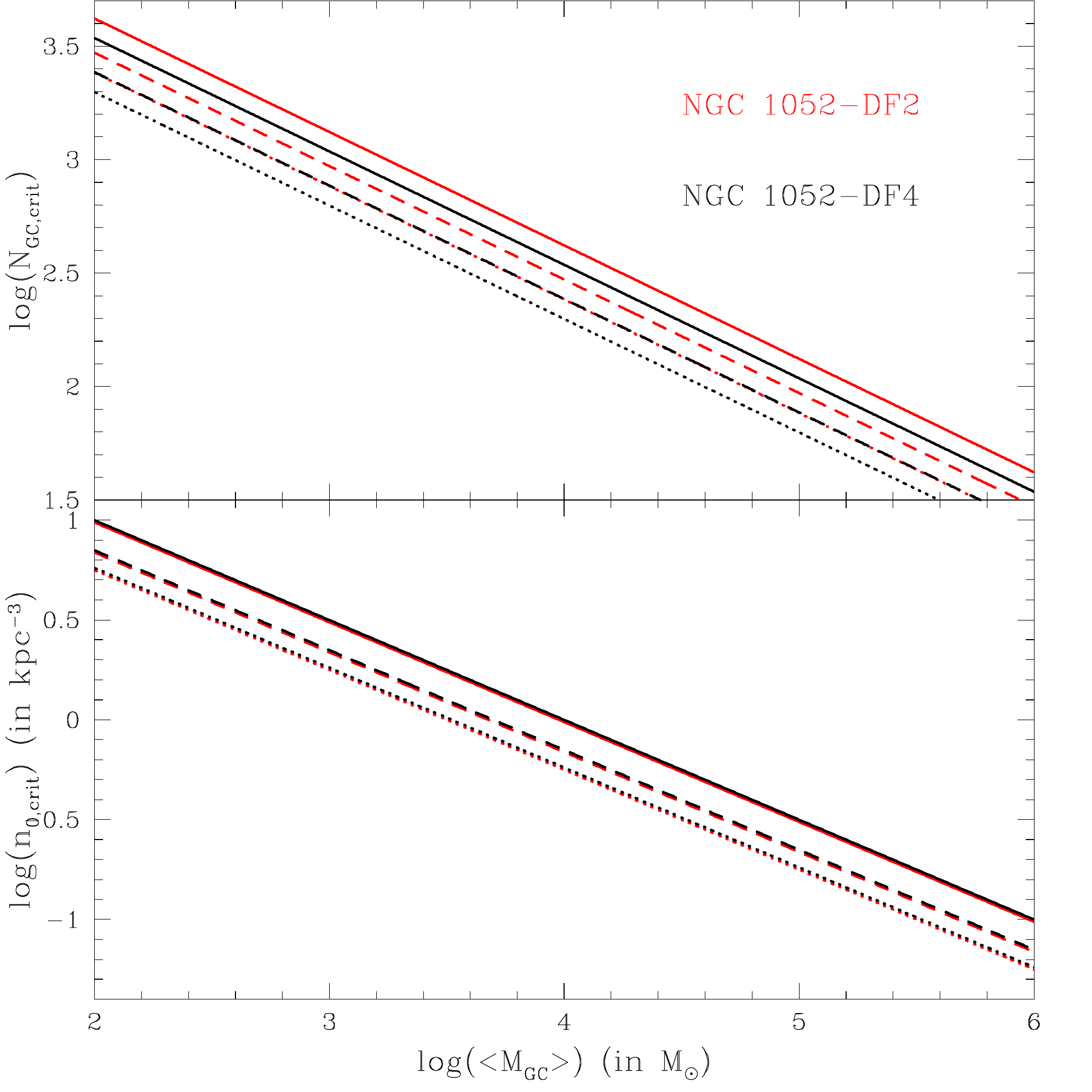}
\end{center}
\caption[The critical number density and critical number of GCs within the specified volumes required for a single GC-GC collision to occur within 1-3 Gyr.]{The critical number density (bottom panel) and critical number (top panel) of GCs within the specified volumes (see text) required for a single GC-GC collision to occur within 1, 2 and 3 Gyr.  The solid, dashed and dotted lines correspond to, respectively, 1, 2 and 3 Gyr.  As before, NGC 1052-DF4 and NGC 1052-DF2 are indicated by, respectively, the black and red lines.
\label{fig:fig5}}
\end{figure}

To better quantify the above, we set the GC-GC collision times equal to 1, 2 and 3 Gyr, chosen somewhat arbitrarily such that of order 10 such collision events would have occurred over a Hubble time.  We then solve for both the critical number density and the critical number of GCs within the above volumes required for a single GC-GC collision to occur within these times.    The result is shown in Figure~\ref{fig:fig5}.  The upper panel of Figure~\ref{fig:fig5} shows that of order $\sim$ 100 GCs are needed in this volume to have a single GC-GC collision within 1 Gyr (see the solid black and red lines).  This implies that, assuming mean GC masses of 2.5 $\times$ 10$^5$ M$_{\odot}$ and 1.3 $\times$ 10$^5$ M$_{\odot}$ for, respectively, NGC 1052-DF2 and NGC 1052-DF4, 10 and 7 GC-GC collisions (i.e., corresponding to the observed number of bright GCs in each galaxy) would happen within 10 Gyr and 7 Gyr, respectively. 

The above calculations show that, had more GCs been present in the past and with a centrally concentrated spatial distribution, this would most likely have resulted in significant collisional evolution.  In turn, the initial properties of the GC populations would have been modified, in particular the observed distributions of GC masses and galactocentric radii.  It is unclear, however, to what degree the GC populations may have been different in the past, and which dynamical histories are viable and allowed, as decided by the need to uphold causality, conservation laws and the underlying physics (e.g., the rate of orbit diffusion in energy- and momentum-space).  As we will show below and in the subsequent section, quantifying and constraining the viable evolutionary channels is one of the main goals of this paper.

To summarize, our results show that, had more GCs been present in the past, some collisions between the most massive GCs would have likely occurred, skewing the observed GC luminosity function to higher GC masses.  As we will show in the next section, apart from this effect which operates to preferentially modify the high-mass end of the GC luminosity function, any collisional evolution of a GC system would contribute to depleting preferentially low-mass GCs via evaporative effects, further skewing the mean of the GC mass function to even higher masses.\footnote{Similarly, we naively expect a prior galaxy-galaxy interaction to preferentially strip low-mass GCs, since the lowest mass orbiters tend to be the most weakly bound.}

\section{The collisional evolution of GC systems}

In this section, we discuss the time evolution of a given GC system due to collisional dynamics, first using a Boltzmann equation for the time evolution of an initial particle mass function and then in the context of Collision Rate Diagrams \citep{leigh17,leigh18}.  We discuss this collisional evolution in the context of a GC system perturbed significantly during a prior galaxy-galaxy interaction, as considered in this paper to explains the origins of the observed properties of the galaxies NGC 1052-DF2 and NGC 1052-DF4.  

\subsection{Quantifying the dynamical evolution of a GC population using a Boltzmann equation}

Ignoring dissipative effects, the dynamical evolution of the GC mass function should be \textit{statistically deterministic}.   Said another way, the fates of individual particles are sensitive to the precise initial conditions, but the evolution of the overall distribution functions are not.  To see that this should indeed be the case, consider the following equation:
\begin{equation}
\label{eqn:boltzmann}
\frac{{\partial}f_{\rm m}}{{\partial}t} + \frac{{\partial}f_{\rm m}}{{\partial}m}\frac{{\partial}m}{{\partial}t} = -\frac{{\partial}}{{\partial}m}\Big( f_{\rm m}<{\Delta}m> \Big) + \frac{1}{2}\frac{{\partial}^2}{{\partial}m^2}\Big( f_{\rm m}<{\Delta}m^2> \Big),
\end{equation}
where f$_{\rm m}$(m) is the GC mass function, which is a continuous differentiable function over the range of GC masses of interest (i.e., from the assumed initial minimum GC mass to the initial maximum GC mass), and $<{\Delta}$m$>$ and $<{\Delta}$m$^2>$ are first- and second-order diffusion coefficients.  The diffusion coefficients can be calculated accordingly:
\begin{equation}
\label{eqn:boltzmann2}
<{\Delta}m> = \int \Gamma(m)f_{\rm m}(m){{\Delta}m}{\rm d}{\Delta}m
\end{equation}
and
\begin{equation}
\label{eqn:boltzmann3}
<{\Delta}m^2> = \int \Gamma(m)f_{\rm m}(m){{\Delta}m^2}d{\Delta}m,
\end{equation}
and $\Gamma$(m) is the mass-dependent collision rate:
\begin{equation}
\label{eqn:boltzmann4}
\Gamma(m) = n(m)\sigma_{\rm coll}(m)v_{\rm rms}(m).
\end{equation}
In the above equation, n(m) and v$_{\rm rms}$(m) are the number density and root-mean-square velocities for GCs with mass m, respectively.  The collisional cross-section is denoted by $\sigma_{\rm coll}$(m), and gives the gravitationally-focused cross-section for collisions involving species of mass m (the total rate for a given mass species can be obtained by integrating the collision rate over the GC mass function).  Both the GC number density and root-mean-square velocity are mass-independent initially \citep{lyndenbell67}, and evolve toward a state of energy equipartition at a rate that can be determined using a multi-mass Fokker-Planck equation (see below).

Equation~\ref{eqn:boltzmann} is a Boltzmann-type of equation that quantifies the evolution of a GC population in mass function-space due to direct collisions.  We are most interested in starting from a well-mixed (in energy-space, and hence position- and velocity-space) population of GCs as occurs post-violent relaxation.  Hence, an episode of violent relaxation provides a well-defined "initial" state (see \citet{lyndenbell67} for more details) from which the subsequent dynamical evolution follows \textit{in a statistically deterministic or causal manner}.  Equation~\ref{eqn:boltzmann} assumes conservation of mass, energy and angular momentum, so does not account for mass loss due to stellar evolution or cluster evaporation in a tidal field, for example.  It can, in principle, be combined with a multi-mass Fokker-Planck model to simultaneously quantify the dynamical evolution of the GC mass function in position- and velocity-space within the host galaxy.  As with the high-mass end, the subsequent evolution of the low-mass end of the GC mass function is statistically deterministic and can be easily parameterized (see \citet{webb15} for more details on how to account for stellar mass loss from individual GCs).

To summarize, the GC mass function will evolve due to two separate effects:  direct collisions quantified by the Boltzmann-type equation in mass-space (i.e., Equation~\ref{eqn:boltzmann}), and the re-distribution of GCs in position- and velocity-space within their host galaxies induced by two-body relaxation quantified via a multi-mass Fokker-Planck 'master equation'.  These two mechanisms preferentially impact, respectively, the high- and low-mass ends of the GC mass function.

\subsection{Quantifying the dynamical evolution of a GC population using Collision Rate Diagrams} \label{crd}

In this section, we present our method for constraining the viable evolutionary channels for the internal dynamical evolution of GC systems in galaxies, given the currently observed GC luminosity functions.  We explain our method using different illustrative examples corresponding either to the limits of very high rates of direct collisions or of impulsive fly-bys.  This is meant to show that the past GC luminosity functions and spatial distributions are uniquely constrained by their present-day observed values.  These constraints are decided by conservation- and diffusion-based arguments which must all be upheld in order to causally connect the subsequent dynamical evolution coupling the final observed states to the allowed set of initial conditions.  Since it is a fundamentally chaotic problem, our method identifies the most \textit{probable} evolutionary pathway, given a set of final observed properties for a given GC system.  More importantly, as we will show, our method forbids large sections of parameter space, ruling these out as possible initial conditions and informing future more sophisticated numerical simulation-based studies.

\begin{figure}
\begin{center}
\includegraphics[width=\columnwidth]{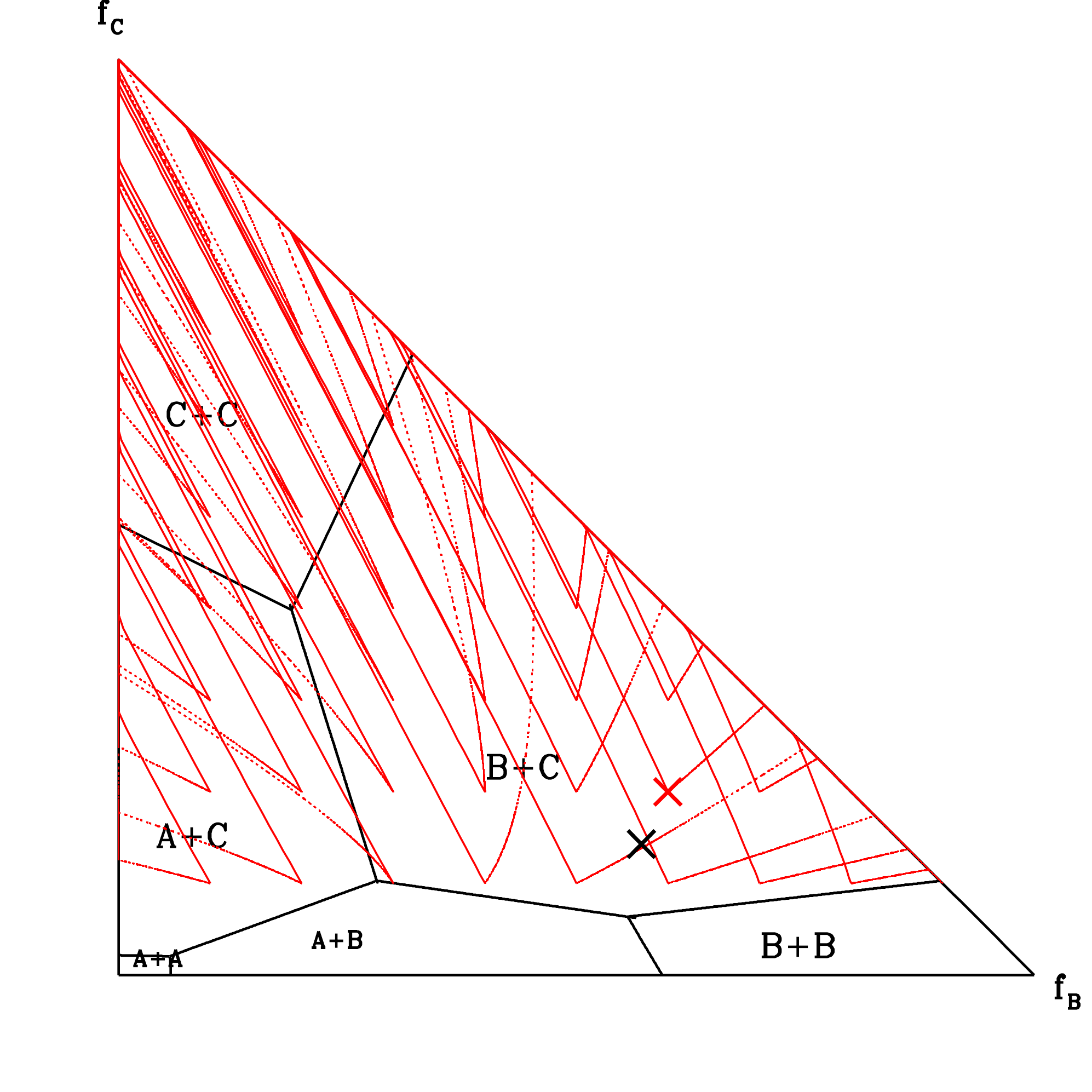}
\end{center}
\caption[Collision Rate Diagram for the GC populations in the NGC 1052 group]{Collision Rate Diagram \citep{leigh17,leigh18} for GC populations in the NGC 1052 group.  We consider three different GC species for a given galaxy, each with its own assumed mass and size, and show the time evolution of their relative number fractions.  We adopt GC types A, B and C and assume that they adhere to a ratio in mass and size of 1:2:$\ge$3, respectively (the units are not relevant for the relative rates, only the absolute rates; see text for more details). Following the procedure described in the text, we plot the fractions of B- and C-type particles on the x- and y-axes respectively, and assume that 1 $=$ f$_{\rm A} +$ f$_{\rm B} +$ f$_{\rm C}$. The different segmented regions indicate where different collision scenarios each dominate. The solid red lines show the time evolution in the f$_{\rm B}$-f$_{\rm C}$-plane for each assumed set of initial relative number fractions.  Finally, the red and black crosses show the observed fractions for, respectively, NGC 1052-DF2 and NGC 1052-DF4 (see text for details).  
\label{fig:fig6}}
\end{figure}

In Figure~\ref{fig:fig6} we show the time evolution of the number fractions of three different GC species.  Each species has an unique combination of mass and radius.  We adopt GC types A, B and C and (conservatively) assume that they adhere to a ratio in mass and size of 1:2:$\ge$ 3, respectively (the units are not relevant for the \textit{relative} rates, only the \textit{absolute} rates).\footnote{Steeper mass ratios would only accelerate the basic trends we report here, rapidly driving host galaxies to very high fractions of only the most massive of their original GCs.  Hence, our assumption here is the most conservative possible, in this regard.}  We then follow the procedure described in \citet{leigh17} and expanded upon in \citet{leigh18} to calculate the relative collision rates for different particles types.  Specifically, we calculate the relative collision rates using Equations 19 and 24 in \citet{leigh17}. We plot the fractions of B- and C-type particles on the x- and y-axes, respectively, and assume 1 $=$ f$_{\rm A} +$ f$_{\rm B} +$ f$_{\rm C}$. The different segmented regions indicate the parameter space where different collision scenarios each dominate.  

As an initial illustrative example, we focus on larger impact parameter interactions than correspond to direct collisions, to quantify the preferential ejection of lower mass GCs.  For simplicity, we assume that interactions between different particle types always eject the lowest-mass GC from the galaxy.  To calculate the time evolution in the relative particle fractions, we assume that interactions between identical particles always result in the ejection of one of the two particles.  These assumptions are over-simplified but capture the general trends expected from simple conservation of linear momentum and energy, given the (mass-independent) initial conditions expected post-violent relaxation (due to, for example, a recent galaxy-galaxy collision \citep{lyndenbell67,madau19}).  We sample the allowed parameter space of initial conditions uniformly in the f$_{\rm B}$-f$_{\rm C}$-plane, and follow the subsequent time evolution in the f$_{\rm B}$-f$_{\rm C}$-plane until only one type of GC remains.  

As is clear from the dotted red lines in Figure~\ref{fig:fig6}, the evolution is always toward very high fractions of C-type particles, which correspond to the most massive GCs given our assumptions.  Within the context of our hypothesis, this is roughly consistent with what is currently observed in NGC 1052-DF2 and NGC 1052-DF4, if the observed GCs represent the remains of once much richer GC populations.  Figure~\ref{fig:fig6} suggests that the present-day observed relative number fractions can be used to uniquely constrain the initial number fractions, since every trajectory (depicted by the red lines) is unique and does not cross any other lines (excluding evolution along the outer boundaries).\footnote{We note that the contribution of internal stellar two-body relaxation within GCs to modifying the observed GC luminosity functions at the preferentially low-mass end is unlikely to change these conclusions.  This is because any subsequent dynamical evolution of the GC systems within these galaxies would only relocate them to parts of their host galaxies where the gravitational potential is lower, reducing the rate of stellar evaporation \citep[e.g.][]{webb15}.  With that said, more extended individual GCs could be more challenging to identify observationally, and this should be taken into account in future observational surveys designed to look for such lower mass GCs in or around these galaxies.}

Another illustrative example is shown in Figure~\ref{fig:fig6}, by the solid red lines.  Here, we focus on smaller impact parameters, leading to direct collisions.  That is, if a close interaction occurs, then so must a direct collision if at the distance of closest approach the stars are closer than the sum of their radii.  We adopt the mass ratios 1:2:$\ge$ 3 corresponding to A:B:C, such that collisions tend to quickly over-populate the C-type particles.\footnote{Note that we indicate $\ge$ 3 here, since we count all collision products as C-type particles, independent of their mass.}  As is clear from this simple exercise, the flow lines in the $f_{\rm B}-f_{\rm C}$-plane are maximally directed toward $f_{\rm C} \sim 1.0$ on short timescales. 

Finally, we apply Figure~\ref{fig:fig6} to the GC populations in both NGC 1052-DF2 and NGC 1052-DF4.  To compute the fractions of particle types, we define A-, B- and C-type particles to correspond to the mass intervals, respectively, 1 - 4 $\times$ 10$^5$ M$_{\odot}$, 4 - 8 $\times$ 10$^5$ M$_{\odot}$ and 8 - 12 $\times$ 10$^5$ M$_{\odot}$.  According to Table~\ref{table:one}, this gives $f_{\rm A} = 0.2$, $f_{\rm B} = 0.6$ and $f_{\rm C} = 0.2$ for NGC 1052-DF2 (red cross) and $f_{\rm A} = 0.3$, $f_{\rm B} = 0.6$ and $f_{\rm C} = 0.1$ for NGC 1052-DF4 (black cross).  

Given the two limiting cases considered above, the positions of the observed data points in Figure~\ref{fig:fig6} suggest that, over time, the relative fractions of B- and C-type particles should have increased.  Thus, we can safely conclude that any internal collisional evolution of these GC systems could only have contributed to further depleting their luminosity functions of preferentially low-mass GCs, with direct collisions/mergers only further skewing the mean toward even higher luminosities.  It follows that the prediction that these galaxies should be depleted in low-mass GCs had prior galaxy-galaxy interactions indeed occurred remains intact independent of any post-interaction dynamical evolution of their GC systems.

The key point to take away from these simple examples is that there is a strong connection between the initial conditions and the present-day observed state of the system. More specifically, from these simple examples, we see that knowing the observed present-day relative GC number fractions and following a deterministic evolutionary path for the subsequent dynamical evolution via collisions and/or ejections, we are able to use the present-day observed number fractions to uniquely constrain the initial relative number fractions prior to any internal dynamical processing.  Thus, the method could potentially allow for the candidate galaxy's \textit{past or pre-dynamically processed} GC luminosity function (and GC specific frequency, etc.) to be uniquely determined.  

This motivates the need to perform additional simulations of strong galaxy-galaxy interactions and their implications for pre-existing GC populations, to help populate Figure~\ref{fig:fig6} and better understand the dependence on the initial conditions and the effects of strong interactions. Our results predict a paucity of low-mass GCs relative to a scenario without any prior host galaxy-galaxy interaction, triggering an episode of violent relaxation and the subsequent internal dynamical evolution of the host galaxy's GC population.  To what degree the assumption of violent relaxation is correct will be quantified in a forthcoming study using more sophisticated numerical simulations, to better understand how energy and angular momentum is transferred to the GC systems of interacting galaxies.  

This also motivates a deeper and more thorough observational campaign to try to identify any additional GCs associated with NGC 1052-DF2 and NGC 1052-DF4, to probe further down the GC luminosity function.  This would allow us to improve our analysis in Figure~\ref{fig:fig6}, and populate it with more robust empirical data.  In turn, this would facilitate more stringent constraints on the conditions post-interaction and, specifically, the initial GC mass function.

\section{Discussion} \label{discussion}

In this paper, we consider the origins of the recently discovered ultra-diffuse DM-free galaxies NGC 1052-DF2 and NGC1052-DF4.  This is because we are presenting a new method for uniquely constraining the past dynamical evolution of GC populations thought to have undergone a prior episode of violent relaxation.  Such an episode of violent relaxation could have occurred due to a prior strong galaxy-galaxy interaction, which has been suggested in the literature as a mechanism to explain the observed properties of the galaxies NGC 1052-DF2 and NGC1052-DF4, hypothesized to be DM-free.  Hence, our method can be used to constrain the viability of such scenarios.  The presently observed properties and numbers of the remaining GCs is all that is required to apply the method robustly.  

In this section, we consider to two different scenarios to account for the observed GC properties, both involving a prior strong interaction with a more massive galaxy that stripped the host galaxies of their DM.  These are:  (1) The progenitors of NGC 1052-DF2 and NGC 1052-DF4 either had no massive GCs prior to the interaction, such that the interaction triggered the formation of the observed massive GCs \citep[e.g.][]{silk19}; or (2) the progenitors had a substantial GC population, and the interaction triggered an episode of violent relaxation in the host galaxy GC population.  In the latter case, violent relaxation would significantly perturb the GC orbits, mixing them thoroughly in phase-space.   This should push some GCs to become highly eccentric, while also ejecting loosely bound GCs.  These ejected GCs would either become "free-floating", or end up gravitationally bound to the more massive interloping galaxy.

As described in Section~\ref{intro}, a close interaction between two galaxies with a large mass ratio between them could strip them of their DM content by ejecting the stars and gas in the progenitor's disk.  This is also the case in the event of a direct high-velocity collision \citep[e.g.][]{silk19}.  During a close interaction, it has been shown \citep[e.g.][]{nusser19b} that, when the DM halo is ejected, the left-over stars and gas comprising the galaxy more or less retain the velocity dispersion of their much more massive progenitor. Similarly, a direct collision will deposit kinetic energy, only some of which is dissipated by the gas.  Thus, immediately after a close interaction or even collision, DM-stripped galaxies should be highly super-virial, and will expand by the virial theorem.  Re-virialization should occur on a crossing time, which for NGC 1052-DF4 happens to currently be $\sim$ 0.2 Gyr within the half-light radius, assuming a half-light radius of 1.6 kpc and a stellar velocity dispersion of 7 km s$^{-1}$ \citep{vandokkum19} (and NGC 1052-DF2 therefore has a similar crossing time, given the similar properties of these two galaxies).  If the progenitor galaxies were initially more compact than is currently observed, then this would only reduce the crossing time, which is already comparable to the shortest DF timescale for our sample of GCs, at least in NGC 1052-DF4. It therefore seems likely that any such stripping event, whether it be a strong close interaction or a direct high-velocity collision, would produce a remnant galaxy that should expand. \textit{This could contribute to, and perhaps even entirely account for, their observed ultra-diffuse state} \citep[e.g.][]{silk19}.

Is it plausible that both NGC 1052-DF2 and NGC1052-DF4 experienced a recent close interaction with another galaxy in the NGC 1052 group?\footnote{Here we note that a direct high-velocity collision would not leave behind an interloping galaxy in the group to search for.}  \citet{vandokkum18a} showed that NGC 1052-DF2 could certainly have recently experienced a close interaction with the most massive galaxy in the group, namely NGC 1052, given their very close proximity in projection. NGC 1052-DF4 lies roughly a factor of two further from NGC 1052 in projection, relative to NGC 1052-DF4. Hence, it is also entirely plausible that it too experienced a close interaction with NGC 1052 in the recent past. For NGC 1052-DF4, however, another galaxy lies even closer to it in projection.  This is NGC 1035, which lies at a projected distance of 23 kpc from NGC 1052-DF4 and has a relative velocity of 204 km s$^{-1}$. Assuming a relative velocity at infinity of 110 km s$^{-1}$ \citep{cohen18}, which is equal to the observed line-of-sight velocity dispersion in the NGC 1052 group, and the minimum possible 3D distance of 23 kpc, we compute an interaction time of only $\sim 0.02$ Gyr, which is roughly an order of magnitude less than our inferred upper limit from the GC DF timescales (see Figure~\ref{fig:fig3}) for the time since a close interaction between NGC 1052-DF4 and another galaxy in the NGC 1052 group must have occurred.  Hence it is entirely feasible, but by no means guaranteed, that NGC 1052-DF4 very recently had a close interaction with NGC 1035.  We conclude that the proposed scenario for stripping both NGC 1052-DF2 and NGC 1052-DF4 of their DM, namely a recent close interaction with a nearby more massive galaxy, is plausible.

\subsection{Implications from different interaction scenarios}

In this section, we consider how the observed properties of NGC 1052-DF2 and NGC 1052-DF4, as well as of their observed GC populations, should change for different interaction scenarios, given our hypothesis of a strong galaxy-galaxy interaction having occurred some time in the past.  We consider two different scenarios:  (1) The progenitors of NGC 1052-DF2 and NGC 1052-DF4 either had no massive GCs prior to the interaction, and the interaction triggered the formation of the observed massive GCs; or (2) the progenitors had substantial GC populations initially, which were significantly perturbed during the interaction, undergoing an episode of violent relaxation \citep{lyndenbell67}.

\subsubsection{Scenario 1:  Did the galaxy-galaxy interaction cause the formation of the observed massive GCs?}
\label{scen1}

In this scenario, we assume that the progenitors of NGC 1052-DF2 and NGC 1052-DF4 recently experienced a strong galaxy-galaxy interaction, and that these hosts initially contained significant mass in gas rotating with their stars.  In this scenario, the more massive galaxy ejects the stars and gas from the DM halo, if the gas is pulled along with the stars, which occurs for comparable orbits (i.e., the spin and orbital frequencies are well-matched, maximizing the magnitude of the effect) and the correct prograde orientation of the interaction.  If the gas were to collect at the bottom of the potential well of the remnant galaxy, the gas densities could become sufficiently high to trigger GC formation in an extreme high-pressure environment, forming more massive GCs at a given cloud density \citep[e.g.][]{murray09,silk19}.  A similar scenario was mostly recently considered by \citet{silk19} who proposed high-velocity fast collisions, which could have produced simultaneous triggering of over-pressurized dense clouds that form preferentially massive globular clusters.  

This scenario immediately predicts stellar ages for the constituents of the GCs observed in NGC 1052-DF2 and NGC 1052-DF4, that are commensurate with the time since the strong interaction occurred.  Hence, naively, this could predict younger (and hence bluer) GCs relative to stars in the field of their host galaxy (and also relative to the GCs described in the below scenario).  Said another way, the minimum DF timescale of all ten/seven GCs in NGC 1052-DF2/NGC 1052-DF4 can be used to put a constraint on the minimum time ago the interactions must have occurred.  For NGC 1052-DF2 and NGC 1052-DF4, these minimum times would be, respectively, $\sim$ 7 Gyr and $\sim$ 1 Gyr.\footnote{But, again, these exact numbers should be taken with a grain of salt, as described previously in the text.} 

We close this section with a brief review of its predictions:

\begin{itemize}

\item The observed massive GCs should be younger than the stars comprising their host galaxy.  Their integrated colours should thus be bluer than that of the host.  If some GCs are retained from the progenitor galaxy, then the GC colour distribution should appear bi-modal.

\item The observed GC luminosity function should be top-heavy and centrally concentrated (expected for GC formation in high-pressure gas-rich environments), with a significant paucity of low-mass GCs.

\item If indeed some GCs have present-day DF timescales that are much shorter than a Hubble time, the shortest of these can be used to constrain the time since the hypothetical galaxy-galaxy interaction or collision occurred, which is needed to rid the host of its DM.  In the case of a very strong close interaction with another perturbing galaxy, this can be converted in to a volume centred on each galaxy within which the more massive perturbing galaxy should reside.

\end{itemize}

\subsubsection{Scenario 2:  Did the galaxy-galaxy interactions significantly perturb pre-existing GC orbits?} \label{scen2}

The expected response of a system of GCs to an episode of violent relaxation is highly collisional, as the system tries to recover a Maxwellian distribution of velocities. That is, the subsequent dynamical evolution is governed by the physics of collisional dynamics, which deterministically connects the initial conditions of the GC populations (i.e., immediately post-violent relaxation) to their final currently observed states.  

If the progenitors of NGC 1052-DF2 and NGC 1052-DF4 both had substantial GC populations before the interaction, the tidal force from the massive perturber would not only perturb them on to highly modified likely eccentric orbits, but also unbind the most tenuously bound GCs.  The subsequent collisional evolution back toward energy equipartition and a Maxwellian distribution of orbital velocities will also eject preferentially low-mass GCs, removing further energy and angular momentum from the most massive GCs and helping to deliver them deeper into the host galaxy potential.  This could predict free-floating GCs somewhere close to NGC 1052-DF2 and NGC 1052-DF4 on the plane of the sky that are not bound to any galaxy.  The interloping massive galaxy (e.g., NGC 1052) could also accrete GCs from the perturbed galaxies, which could be identified if significant age, chemical, etc. differences happen to exist between the native and accreted GCs.  However, if the interaction happened sufficiently far in the past, any free-floating GCs would have had sufficient travel time to have become difficult, if not impossible, to identify observationally.

Is it possible that the observed GCs in both NGC 1052-DF2 and NGC 1052-DF4 began much further out in their host galaxy potential, and have simply been caught in the act of spiraling inward due to DF?  This was recently proposed by \citet{chowdhury19}, who use a suite of 50 multi-GC N-body models to follow the orbital decay of the GCs.  They find that over $\sim$ 10 Gyr many GCs experience significant orbital decay due to DF, whereas others evolve much less.  In their simulations, they find that a combination of reduced DF in the galaxy core and GC-GC scattering keeps the GCs buoyant in their host galaxy potential, such that they have not yet sunk to its centre.  The authors conclude that if NGC 1052-DF2 is indeed devoid of DM, then at least some of its GCs must have formed further out before spiraling in to their current locations, and that the GC system was likely more extended in the past.  \citet{nusser19} used a similar approach to study DF in NGC 1052-DF2 using N-body simulations, and found much the same thing, but with the added correction that in some simulation realizations GCs do decay all the way to the centre of their host galaxy.  Both of the conclusions arrived at in these papers via more detailed N-body simulations are consistent with the overall results reported in this paper.

If \citet{vandokkum19} indeed caught one out of seven GCs in NGC 1052-DF4 at the end of its spiral-in phase (see Figure~\ref{fig:fig3}), then why have no other GCs already spiraled in to the nucleus?  Provided the true DF timescales for these three GCs are close to our calculations, the lack of a central NSC is indeed puzzling.  If other DM-free galaxies are identified in the NGC 1052 group (or any other), the probability that they will host a central NSC could be high, produced by DF of GCs formed or perturbed onto orbits deeper in the host galaxy potential during the close galaxy-galaxy interaction presumed to have stripped its host of its DM.  

Indeed, the observational results of \citet{graham09}, comparing super-massive black hole (SMBH) and NSC masses as a function of their host galaxy mass, suggest that both NGC 1052-DF2 and NGC 1052-DF4 are of sufficiently low mass that their central regions should be dominated by an NSC (if a central massive object, either NSC or SMBH, is present at all), rather than an SMBH.  And yet, close inspection of Figure 1 in \citet{vandokkum19} suggests that no central nuclear star cluster (NSC) is present in either NGC 1052-DF2 or NGC 1052-DF4.  Alternatively, the lack of a central NSC could be pointing toward a stalling of DF as GCs reach the centre of their host galaxy, either due to GC-GC interactions, the inclusion of a radial-dependence to the Coulomb logarithm, etc. (see \citet{chowdhury19} for more details).   

We close this section with a brief review of its predictions:

\begin{itemize}

\item Relative to galaxies where GC-GC collisions are not expected to happen, this predicts a top-heavy GC mass function, with the brightest and hence most massive GCs residing at small galactocentric distances.  This last effect should be enhanced via the fact that direct GC-GC collisions dissipate both orbital energy and angular momentum, causing the collision products to fall even deeper in to the host galaxy potential.

\item The observed GCs should have roughly the same age as the stars comprising their host galaxy.  Their integrated colours, corresponding to old stellar populations, should thus be very similar to that of their host.

\item The distribution of (3D) GC velocities should be close to isotropic.

\item The most tenuously bound GCs in the galaxy progenitors could have been stripped or accreted on to the more massive interloping galaxy (e.g., NGC 1052).  This could predict non-native GCs in the (hypothetical) more massive perturbing galaxy that were accreted during the interaction.  Alternatively, it could predict free-floating GCs lingering as debris in the vicinity of each galaxy post-interaction.  The observability of such free-floating GCs is, however, likely to be very sensitive to exactly \textit{when} the hypothetical galaxy-galaxy interaction occurred.  If our computed DF timescales are accurate, identifying these free-floating galaxies should be most probable for NGC 1052-DF4, given its much shorter constraint on the time since the interaction occurred. That is, any free-floating GCs produced would only have a travel time of $\sim$ 1 Gyr. 

\end{itemize}

\section{Summary}
\label{summary}

In this paper, we present a new method for uniquely constraining the past dynamical evolution of GC populations thought to have undergone a past episode of violent relaxation.  The presently observed properties and numbers of the remaining GCs are all that is required to apply the method robustly.  We consider two different scenarios to account for the observed GC properties, both involving a prior strong interaction with a more massive galaxy.  The encounter is hypothesized to have both stripped NGC 1052-DF2/NGC 1052-DF4 of their DM halos and either triggered their formation or an episode of violent relaxation in the progenitor GC population. 

We first consider the currently observed state of the GC populations in NGC 1052-DF2 and NGC 1052-DF4, from which we infer and quantify the implications for their past and future states.   We calculate the DF timescales for infall to the central nucleus for the GCs in both galaxies.  We find that two out of ten GCs in NGC 1052-DF2 and one out of seven in NGC 1052-DF4 have DF timescales less than a Hubble time.  In principle, the shortest DF time should put a limit on the time since any past galaxy-galaxy interaction occurred.  

For each galaxy, we go on to calculate the critical number of GCs and the critical GC number density needed for a given number of direct GC-GC interactions/collisions to have occurred since the hypothesized galaxy-galaxy interaction.  This is done by setting the number of collisions equal to the observed numbers of bright GCs in each galaxy, and requiring that the evolution occur on a timescale shorter than the minimum DF time in each galaxy.  The results of this analysis show that significant collisional evolution of a richer GC population than is currently observed could have feasibly evolved dynamically to produce the currently observed distributions of GC masses and galactocentric radii.  As described below, this would contribute both to the observed top-heavy GC mass functions and their centrally concentrated galactocentric distances, and motivates more detailed N-body simulations in future work. 

We further present a novel method to constrain the initial GC mass functions prior to the (hypothesized) chaotic dynamical evolution that should occur post-galaxy-galaxy interaction.  To this end, we apply a Collision Rate Diagram to re-wind the clock and constrain the relative numbers of GCs in different mass bins at the time of interaction.  As described in more detail below, this simple exercise motivates obtaining more complete observations of the GC luminosity functions in these galaxies, which can then be used to constrain the origins of the hypothesized DM-free galaxies, by combining the method presented here with a suite of numerical simulations.  

Our key results can be summarized as follows.  For the GC luminosity functions in these galaxies, our results show that a previous galaxy-galaxy interaction could explain any observed lack of low-mass GCs once deeper observations have been performed, whereas no previous galaxy-galaxy interaction predicts that many more low-mass GCs should be found.  For the GC spatial distributions, our results suggest that a previous galaxy-galaxy interaction could explain a diffuse spatial distribution and/or a paucity of low-mass GCs given the centrally concentrated distribution of the high-mass observed GCs, whereas no previous galaxy-galaxy interaction predicts that many more low-mass GCs should be found with a spatial distribution similar to what is observed for the high-mass GCs \citep[e.g.][]{vandokkum18a,vandokkum19,trujillo19}.  Our results further show that, by adding only a few more GCs in the past, some collisions between the most massive GCs would have likely occurred, further skewing the observed GC luminosity function to higher GC masses.

\section*{Acknowledgments}

We thank Adi Nusser, Asher Wassermann and Pieter van Dokkum for their insightful comments on an early version of the paper. We are grateful to Asher Wasserman for providing data of the star clusters, and Doug Geisler for useful discussions. NWCL is gratefully supported by a Fondecyt Iniciacion grant (11180005).  GF is supported by the Foreign Postdoctoral Fellowship Program of the Israel Academy of Sciences and Humanities. GF also acknowledges support from an Arskin postdoctoral fellowship.

\label{lastpage}

\end{document}